\documentclass{article}

\usepackage{arxiv}
\usepackage{amsmath}
\usepackage[utf8]{inputenc} 
\usepackage[T1]{fontenc}    
\usepackage{hyperref}       
\usepackage{url}            
\usepackage{booktabs}       
\usepackage{amsfonts}       
\usepackage{nicefrac}       
\usepackage{microtype}      
\usepackage{lipsum}		
\usepackage{graphicx}
\usepackage{natbib}
\usepackage{doi}

\title{First analysis of  \textit{in-situ} observation of surface Alfv\'en waves in ICME flux rope}

\author{Anil Raghav${1^*}$, Omkar Dhamane$^1$, Zubair Shaikh$^2$, Naba Azmi$^1$,Ankita Manjrekar$^{1,2}$,\\ Utsav Panchal$^1$, Kalpesh Ghag$^1$, Daniele Telloni$^3$, Raffaella D'Amicis$^4$, Prathmesh Tari$^1$ Akshata Gurav$^1$ \\
 \\
 $^1$Department of Physics, University of Mumbai, Mumbai, India\\
 $^2$Indian Institute of geomagnetism, Panvel, Navi Mumbai, India\\
 $^3$National Institute for Astrophysics, Astrophysical Observatory of Torino, Via Osservatorio 20, I-10025 Pino Torinese, Italy\\
  $^4$National Institute for Astrophysics, Institute for Space Astrophysics and Planetology,\\ Via del Fosso del Cavaliere 100, I-00133 Roma, Italy \\
	\texttt{*anil.raghav@physics.mu.ac.in} \\
   }

\renewcommand{\shorttitle}{Raghav et. al. }

\hypersetup{
pdftitle={A template for the arxiv style},
pdfsubject={q-bio.NC, q-bio.QM},
pdfauthor={David S.~Hippocampus, Elias D.~Striatum},
pdfkeywords={First keyword, Second keyword, More},
}

\begin{document}
\maketitle
\begin{abstract}
		Alfv\'en waves (AWs) are inevitable in space and astrophysical plasma. Their crucial role in various physical processes has triggered intense research in solar-terrestrial physics. Simulation studies have proposed the generation of AWs along the surface of a cylindrical flux rope, referred to as Surface AWs (SAWs); however the observational verification of this distinct wave has been elusive to date. We report the first \textit{in-situ} observation of SAWs in an interplanetary coronal mass ejection flux rope. We apply the Wal\'en test to identify them. We have used Elsa\"sser variables to estimate the characteristics of SAWs. They may be excited by the movement of the flux rope's foot points or by instabilities along the plasma magnetic cloud's boundaries. Here, the change in plasma density or field strength in the surface-aligned magnetic field may trigger SAWs.
\end{abstract}


\section{Introduction}
	The magneto-hydrodynamic  Alfv\'en waves (AWs) are ubiquitous plasma wave modes in space and astrophysical regimes. In these waves, ions collectively respond to perturbations in the ambient magnetic field, such that the ions provide inertia, while the magnetic field supplies the required restoring force \citep{alfven1942existence}. The fluid velocity and magnetic field fluctuations propagating along the magnetic tension force, i.e. well-correlated changes in the respective components of the magnetic field and plasma velocity leads to the apparent characterization of AWs  \citep{walen1944theory,hudson1971rotational,yang2013alfven,raghav2018first}.
	In heliospheric plasma, AWs are observed in two forms: arc-polarized Alfv\'en waves that have often been recognized in the solar wind \citep{belcher1971large,wang2012large} and tube modes in ideal magnetic ﬂux ropes, such as the torsional mode \citep{gosling2010torsional,raghav2018torsional}. These modes are appealing as they carry significant energy from the subphotospheric regions to the corona, and provide energy for coronal heating \citep{van2008detection}. There is a high possibility of identifying AWs in interplanetary space when a magnetic flux rope erupts, no matter the mode in which it is present \citep{wang2019multispacecraft}, However sometimes they are hardly distinguishable from a flux rope configuration \citep{higginson2018structured}, therefore their interrelationships are more complex than several reported studies.

It is worth to note that the coronal mass ejection (CME) is a eruption of enormous energy and massive magnetized plasma from the solar corona into the heliosphere in form of magnetic flux rope \citep{webb2012coronal,howard2011coronal}. Magnetic reconnection or catastrophe processes are expected to trigger low-frequency AWs, and fast \& slow-mode magnetoacoustic waves during the initiation of CME \citep{kopp1976magnetic}. Thus, the Sun is considered a significant source of outward AWs (\citep{belcher1971large}). Moreover, the reported inward AWs suggest different generation mechanisms apart from the ones mentioned above. Inward AWs are observed in  (1) regions of back-streaming ions from the Earth's bow shock \citep{he2015sunward}, and (2) immediately upstream and downstream regions of reverse shocks associated with corotating interaction regions or interplanetary counter part of CMEs (ICMEs)\citep{tsurutani2009magnetic}. AWs are also found in the vicinity of reconnection exhausts and during the drifting of sunward proton beams in the solar wind \citep{belcher1971large,roberts1987origin,bavassano1989large,gosling2009one,gosling2011pulsed,gosling2009one,gosling2011pulsed}. It is also proposed that it may be triggered by some physical processes happening locally \citep{bavassano2001radial,bruno2013solar}.  Recently, we found their existence during the CME-CME and CME-HSS interactions, and inside the ICME sheath regions \citep{raghav2018first,raghav2018does,dhamane2022observation,raghav2022situ}. Moreover, AWs play a key role in modulating the recovery phase of geomagnetic storms and slowing down the restoration of the magnetosphere toward its pre-storm equilibrium state \citep{raghav2018torsional,shaikh2019concurrent,raghav2019cause, choraghe2021properties,telloni2021alfvenicity}. 	
	
%

The most interesting MHD surface wave in the astrophysical domain is the Surface Alfv\'en  Wave (SAW). It forms when there's a finite thickness boundary between two regions of plasma with substantial inhomogeneity in a magnetic field and/or density \citep{evans2009surface}. SAWs may propagate through surface and filamentary structures (e.g., discontinuities) in interplanetary and interstellar space \citep{wentzel1979hydromagnetic}. 
Their coupling with kinetic Alfv\'en waves dissipate them and heat up the surface  \citep{chen1974theory,hasegawa1976kinetic}.
Moreover, \cite{wentzel1979hydromagnetic} suggested that the SAWs may be triggered by the footpoints movement of the flux tubes, or instabilities along the plasma boundaries. Theory and experiment suggest that the SAW eigenmodes play a crucial role in the AWs heating process \citep{ruderman1996unified, amagishi1986experimental}. The simulation study of the collision between a shock wave and a magnetic flux tube shows that SAWs can be generated and propagated along the flux tube  \citep{sakai2000simulation}. They further suggest that SAWs are possible for magnetic flux tubes with weak electric current, whereas body AWs may be generated when the current is strong. Moreover, \citep{lehane1972propagation} confirmed the SAWs in the laboratory. 

\cite{alfven1942existence} investigated the properties of plasma, assuming the plasma medium to be a highly conducting, magnetized and incompressible fluid. He found that a distinctive wave mode arises in the fluid, propagating along the magnetic field direction known as shear or torsional Alfv\'en wave \citep{cramer2011physics}. It propagates energy through the medium's intrinsic elastic and tensile characteristics. Whereas, Surface Alfv\'en  waves are a second class of waves that are supported by the presence of nonuniformities, such as variations in the Alfv\'en speed. They become more significant as the scale length of the nonuniformity diminishes \citep{ionson1978resonant}.  
Unlike ordinary hydromagnetic body waves,  surface Alfv\'en waves are supported by the elasticity afforded by nonuniformities. 
In the solar wind, surface waves can do work on the expanding gas of the wind much as ordinary body hydromagnetic do \citep{hollweg1975waves}. In an infinite uniform plasma,  Alfv\'en waves are driven solely by the magnetic tension force and that they are the only waves that propagate vorticity. The displacements are vortical and incompressible. \cite{goossens2012surface} investigated  modification of MHD waves in presence of nonuniform density and/or magnetic field. They find that the incompressible surface Alfv\'en waves have vorticity equal to zero everywhere except at the discontinuity, where all vorticity is concentrated. Thus,  the behavior of the surface Alfv\'en waves is clearly different from that of the classic Alfv\'en waves.

Theory and simulation studies suggest the existence of SAWs in magnetic flux rope structures, but observation evidences has yet to be found either in small- or in large-scale (like the ICME) flux ropes. Here, we investigated 401 ICME events listed in the Earth directed ICME catalog measured by WIND Spacecraft and hunted for the \textit{in-situ} evidences of SAWs. Interestingly, we found 3 potential events for SAWs. Out of them, the best event is discussed here in detail. To the best of our knowledge, this is the first observational report of the SAWs superposed on the ICME flux rope.

\begin{figure}
	\centering
	\includegraphics[width = \columnwidth]{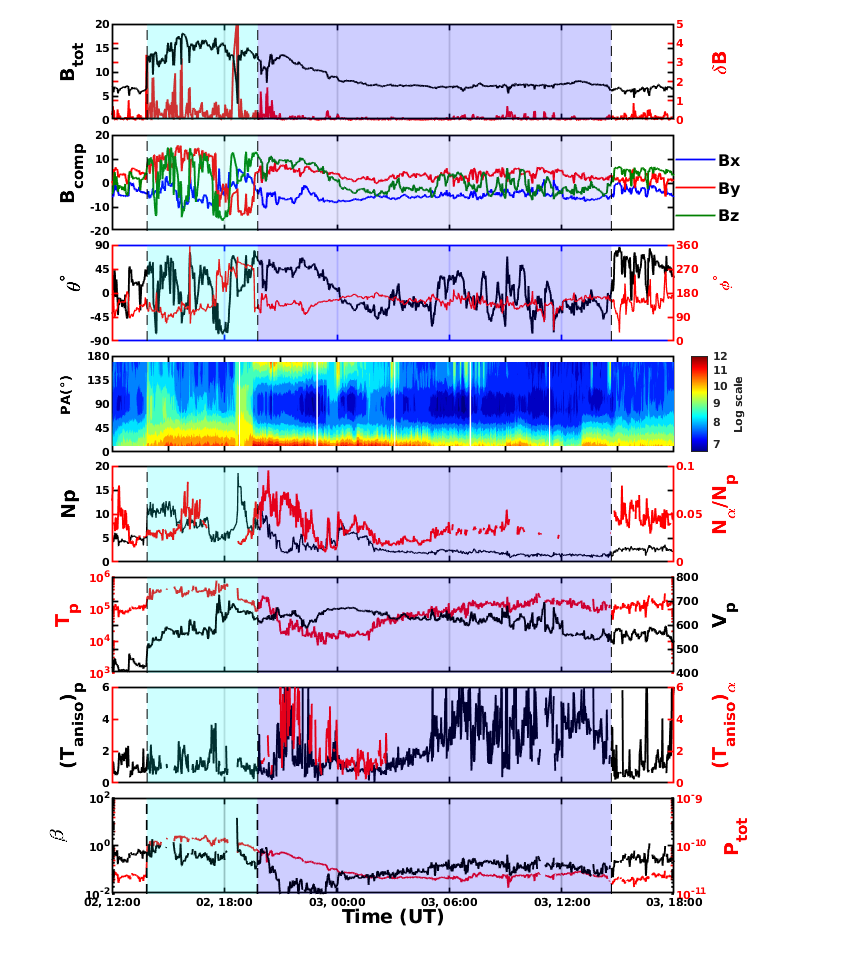}
	\caption{The WIND data over the interval 2005 September 2 12:00:00 to 2005 September 3 18:00:00 UT are plotted. From top to bottom, Interplanetary Magnetic Field (IMF) strength  $B_{tot}$ overlaid with  $\delta B$, a component of the magnetic field $B_{comp}$, the IMF orientation of $\theta$ and $\phi$, the pitch angle (PA) of suprathermal electron strahls, the variation of number density $N_p$ and $\frac{N_\alpha}{N_p}$. variation of temperature of proton ($T_p$) along with velocity of proton ($V_p$), anisotropy in proton temperature $(T_{aniso})_p$ and $\alpha$ particle $(T_{aniso})_{\alpha}$, plasma $\beta$ overlaid with plasma thermal pressure $P_{tot}$ are plotted. }
	\label{fig:IP_label}
\end{figure}

The ICME flux-rope event was identified on 02 September, 2005 by the WIND spacecraft. 
We have used magnetic field and plasma data (of 3 sec time resolution) from Magnetic Field Investigation (MFI) \citep{lepping1995wind} and 3DP Experiment (SWE) \citep{ogilvie1995swe} instruments onboard the WIND spacecraft in Geocentric solar ecliptic (GSE) coordinates.
Their variation with 92-second time resolution during the studied event is demonstrated in figure ~\ref{fig:IP_label}.  The sudden increase in the total magnetic field $B_{tot}$, number density $(N_p)$, total pressure $(P_{tot})$, and plasma temperature $(T_p)$ indicates the onset of the shock front. The Rankine–Hugoniot relation confirms the presence of the shock. The details are available at CfA Interplanetary Shock Database, i.e., \url{https://www.cfa.harvard.edu/shocks/wi_data/00530/wi_00530.html}. The high fluctuations are observed in each magnetic field components, and $\delta B$ in the following region is generally referred to as ICME sheath region (see cyan shaded interval). We observe $\beta$ to be valued near one along with high proton density, plasma temperature and enhanced solar wind speed in this region. The ICME flux rope follows the sheath region (see blue shaded region). ICME flux-rope shows gradual decrease in total magnetic field, decrease in The fluctuations in magnetic field components, low plasma temperature, and low plasma $\beta$ value. Moreover, during flux-rope transit, we observed a nearly bidirectional flow of electron in electron pitch angle variation.  
The different boundaries are defined in two distinct catalogs available online, i.e., Richardson/Cane ICMEs catalog available at \url{http://www.srl.caltech.edu/ACE/ASC/DATA/level3/icmetable2.htm} and  USTC ICME catalog available at \url{http://space.ustc.edu.cn/dreams/wind_icmes}.  

\begin{figure}
	\centering
	\includegraphics[width = \columnwidth]{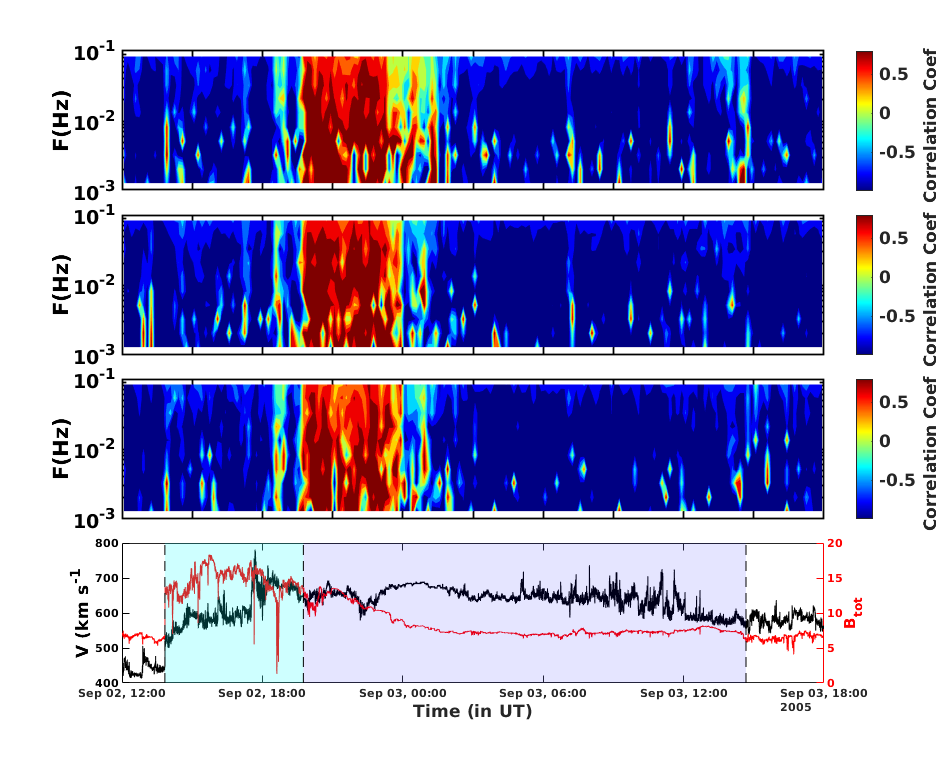}
	\caption{The time-frequency domain gives the correlation coefficient between $V_{Ai}$ and $V_i$ for the entire event. The bottom panel displays the change in the magnetic field and proton velocity over time. The vertical dashed line shows the boundaries of the magnetic cloud and shock sheath of the ICME. For the above analysis, we use 3-second data from the WIND spacecraft.}
	\label{fig:filter}
\end{figure}

To identify Alfv\'en wave, we have employed the Wal\'en relation described as \citep{hudson1971rotational,walen1949vibratory}

\begin{equation*}
	\Delta V ~= ~|R_{W}| ~\Delta V_{A}
\end{equation*}

Here, the linear relation between fluctuations in Alfv\'en velocity ($\Delta V_A = \frac{\Delta B}{\sqrt{\rho_p \mu_0}}$) and solar wind velocity ($\Delta V$ ) provides the Wal\'{e}n slope ($R_w$).
The fluctuations in the magnetic field $\Delta B$ and proton flow velocity $\Delta V$ are determined by removing the background field value ($B_0$) from measured values. The significant correlation between each respective component of $\Delta V_A$ and $\Delta V$ indicates the presence of Alfv\'{e}n wave.  Determining the background value of $V_A$ and $V$ is challenging while implementing the Wal\'en relation. Reported studies have either used the average values of $V_A$, and $V$ of the studied duration, or the values derived by de Hoffmann-Teller (HT) frame \citep{gosling2010torsional,yang2013alfven,raghav2018first,raghav2018torsional,raghav2018does, shaikh2019coexistence}. Here, we used the $4^{th}$ ordered Butter-worth band-pass-filter MATLAB-based algorithm to estimate background values. The evenly divided ten logarithmic frequency bands were selected. The selected bandpass periods are 10s-15s, 15s-25s, 25s-40s, 40s-60s, 60s-100s, 100s-160s, 160s-250s, 250s-400s, 400s-630s, and 630s-1000s. The complete data under examination is split into windows of 200 data points, i.e. 10-minute time window. For every window and filtered band, we find the correlation coefficient between the respective components of $V_A$ and $V$. Figure ~\ref{fig:filter} describes the contour plot of $V_{Ai}$ and $V_i$ along with the temporal variations of total magnetic field and velocity. The sheath and trailing edge of the Magnetic Cloud (MC) indicate a strong negative correlation coefficient (dark blue shade). It confirms that these regions are superposed with the Anti-sunward flow of AWs. The leading part of the MC exhibits a strong positive correlation coefficient (red shade), confirming the sunward propagation of AWs. 

\begin{figure}
\centering
\hspace*{-1.05cm}
\includegraphics[width = 12.5cm]{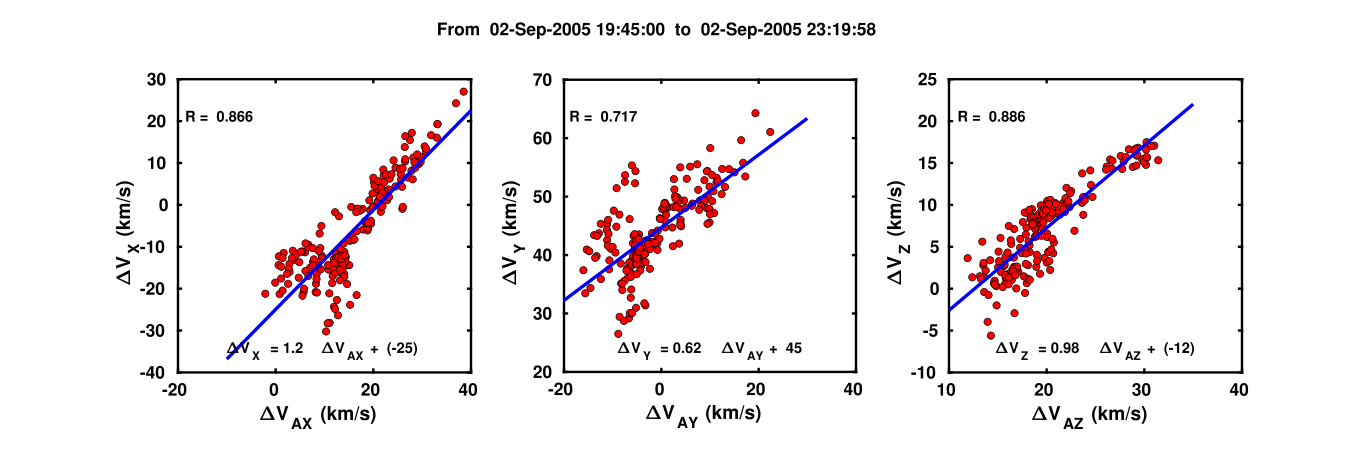}
\hspace*{-1.05cm}
\includegraphics[width = 10.5cm]{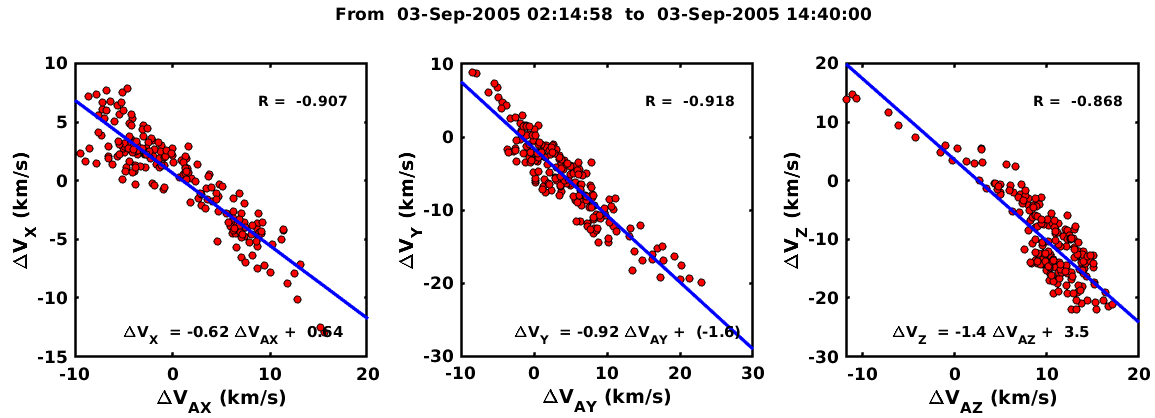}
\caption{Plot shows the analysis of the correlation between the corresponding $\Delta V$ and $\Delta V_A$ components. The scattered black circle with filled red colour represents the WIND spacecraft observations with a time cadence of 3s. The coefficient of correlation is denoted by R. Each panel's equation represents the linear fit relationship between the respective components of $\Delta V$ and $\Delta V_A$.}
\label{fig:cloud_label}
\end{figure}

To support our findings, we estimate correlation coefficients between $\Delta V$ and $\Delta V_A$ for front part of the MC and rear part of MC which depicts in  figure \ref*{fig:cloud_label}.  We used the $4^{th}$ order butter-worth MATLAB filter algorithm with a single broadband frequency boundary of $10^{-3}$ to $10^{-1}$ Hz to filter the $\Delta V$ data and $\Delta V_A$ components.  The Pearson correlation coefficients (R) for each x, y, and z component for the ICME (MC Start time - 02-sep-2005 19:45:00 to 02-sep-2005 23:19:58 ) are 0.866, 0.717, and 0.886, respectively. The strong positive correlation confirmed the sunward nature of the Alfv\'en waves present in the aforesaid region.  Similarly for 03-sep-2005 02:14:58 to 03-sep-2005 14:40:00 we found the Pearson correlation coefficients (R) for each x, y, and z component as -0.907, -0.918, and -0.868, respectively. Hence this negative correlations concludes the presence of strong anti-sunward Alfv\'en waves.   


	\begin{figure}
		\centering
		\includegraphics[width = \columnwidth]{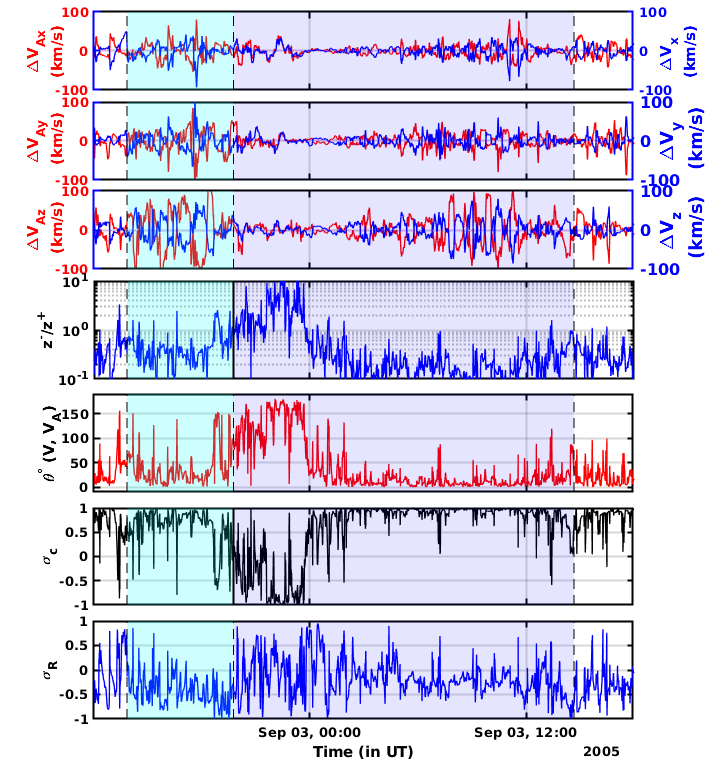}
		\caption{The top three panel shows the temporal variation of Alfv\'en velocity fluctuation $\Delta V_{Ai}$ (red) with proton flow velocity fluctuations $\Delta V_{i}$ (blue). Ratio of Elsa\"sser variables ${z^-}/{z^+}$ shown in the fourth panel. The presence of the angle between the magnetic field and solar wind speed is shown in the fifth panel. The last two panels represent the temporal variation of the normalized cross helicity ($\sigma_c$) and normalized residual energy ($\sigma_R$), respectively.}
		\label{fig:my_label}
	\end{figure}

Figure ~\ref{fig:my_label} depicts the plasma properties in the ICME's shock sheath and MC region based on  Elsa\"sser variables. The top three panels of Figure ~\ref{fig:my_label} show the fluctuations in  Alfv\'enic velocity ($\Delta V_{A}$) and proton flow velocity  ($\Delta V_p$) components, respectively. The fluctuations in each component are obtained by passing each measured component through  $4^{th}$ order butter-worth filter (with frequency limits of $10^{-3}$ to $10^{-1} Hz$ ) algorithm of MATLAB software. We observed anti-correlated flow fluctuations in the sheath region for each component. Interestingly, we found correlated fluctuations flow at the initial part of the MC, whereas an anti-correlated flow is seen in the trailing part of the MC. Here, we employ Els\"asser variables to separate the contributions of outward and inward Alfv\'enic fluctuation flow. The Elsa\"sser variables are defined as \citep{dobrowolny1980fully,zhou1989non,marsch1987ideal,elsasser1950hydromagnetic};
	
	\begin{equation} \label{zpm}
			\Vec{z} ^{\pm} = \Delta \Vec{V}\pm \frac{\Delta \Vec{B}}{\sqrt{4\pi\rho}} = \Delta \Vec{V}\pm\Delta \Vec{V_A}
	\end{equation}

Here, Els$\ddot{a}$sser variables $\vec{z}^+$ and $\vec{z}^-$ are set up to find the waves flow direction, i.e., outward and inward, respectively \citep{roberts1987origin}.
The $\pm$ sign in front of $\vec{B}$ depends on the sign of [$-k ~\cdotp B_0$].
We observe that, for the ratio (${z^-}/{z^+}$) $< 1$ the outward flow became more effective in the sheath region and trailing part of the MC, whereas for the ratio $> 1$ the inward  Alfv\'enic fluctuations are dominant in the initial part of MC    \citep{matthaeus1982stationarity,tu1989basic}. The  angle  between  Alfv\'{e}n velocity  and  solar wind velocity $\theta(V,V_A)$ is estimated as;
\begin{equation*}
	\theta (\Delta V_A ,\Delta V)=arccos(\frac{\Vec{\Delta V_A.}\Vec{\Delta V} }{||\Vec{\Delta V_A}||~||\Vec{\Delta V}||})
\end{equation*}
At the beginning of the MC, both vectors are almost anti-parallel to each other, while in the sheath region and trailing part of the MC, the flow is concurrent. This strongly suggests flow direction changes in each region.



 Cross helicity $(H_c)$ demonstrates a measure of the correlation between velocity and magnetic field \citep{bruno2013solar},  the dimensionless measure of cross helicity is known as normalized cross helicity ($\sigma_c =\frac{H_c}{E})$ ranging from $-1 $ to 1. For $\sigma_c =\pm1$, the fluctuations are highly Alfv\'enic. 
  \begin{equation} 
 	\sigma_c = \frac{{e^+}-{e^-}}{{e^+}+{e^-}}
 \end{equation}
Here $e^-$ $\&$  $e^+$ are the energies related to $z^-$ and   $z^+$ and $e^{\pm}$=$\frac{1}{2}$ $(z^{{\pm}})^2$. Also, the  normalized residual energy is calculated as \citep{bruno2013solar,bruno2005observations}
\begin{equation}
	\sigma_R=\frac{e^v-e^b}{e^v+e^b}
\end{equation}
where $e^v$ $\&$ $e^b$ is kinetic and magnetic energy respectively. We observe $\sigma_c$  $> 0$ , $\sigma_R$  $< 0$ in sheath region and trailing part of the MC(outward). Also, $\sigma_c$  $< 0$, $\sigma_R$  $> 0$ in the leading part of MC(inward). These observations strongly agree with the estimation of Els$\ddot{a}$sser variables.
 	 
 

\begin{figure}
		\centering
		\includegraphics[width = 1 \columnwidth]{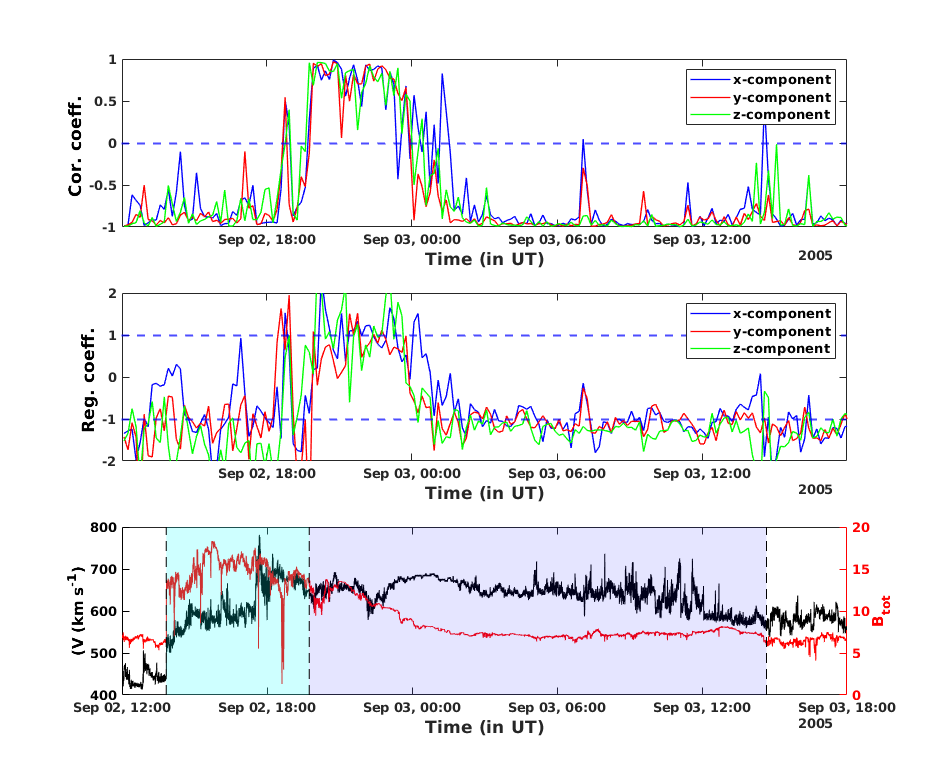}
		\caption{The correlation $\&$ regression coefficient is shown in the top two panels, while the variation of Total magnetic field($B_{tot}$) \& solar wind speed (V) with time is shown in the bottom panel.}
		\label{fig:RC}
\end{figure}

The intensity of inward-outward propagating waves or their mixing is investigated by using the temporal variations of Wal\'en slope (or the correlation between the magnetic field and plasma velocity) \citep{shiota2017turbulent,yang2016observational,belcher1971large,marsch1993modeling,bruno2013url}. 
Their temporal fluctuations are depicted in figure ~\ref{fig:RC}. Both coefficients fluctuate near the value of $\sim -1$ in the sheath region and trailing part of the MC, whereas they fluctuate $\sim 1$ in the leading part of the MC.

All the observations and estimations suggest an outward Alfv\'en fluctuation flow in the sheath region, an inward flow in the front part of MC, and an outward flow in their end part. It implies two possibilities for an AW generation; (i) the interaction between the sheath and MC triggers an oppositely directed wave flow in the sheath region and MC or (ii) the change in the axial current induces Alfv\'enic fluctuations in the magnetic flux rope. The simulation study corroborates that the collision between a shock wave and a magnetic flux tube describes SAWs generation, or the possibilities of SAWs for magnetic flux tubes with weak electric current \citep{sakai2000simulation}.

\begin{figure}
	\centering
	\includegraphics[width = 0.85\columnwidth]{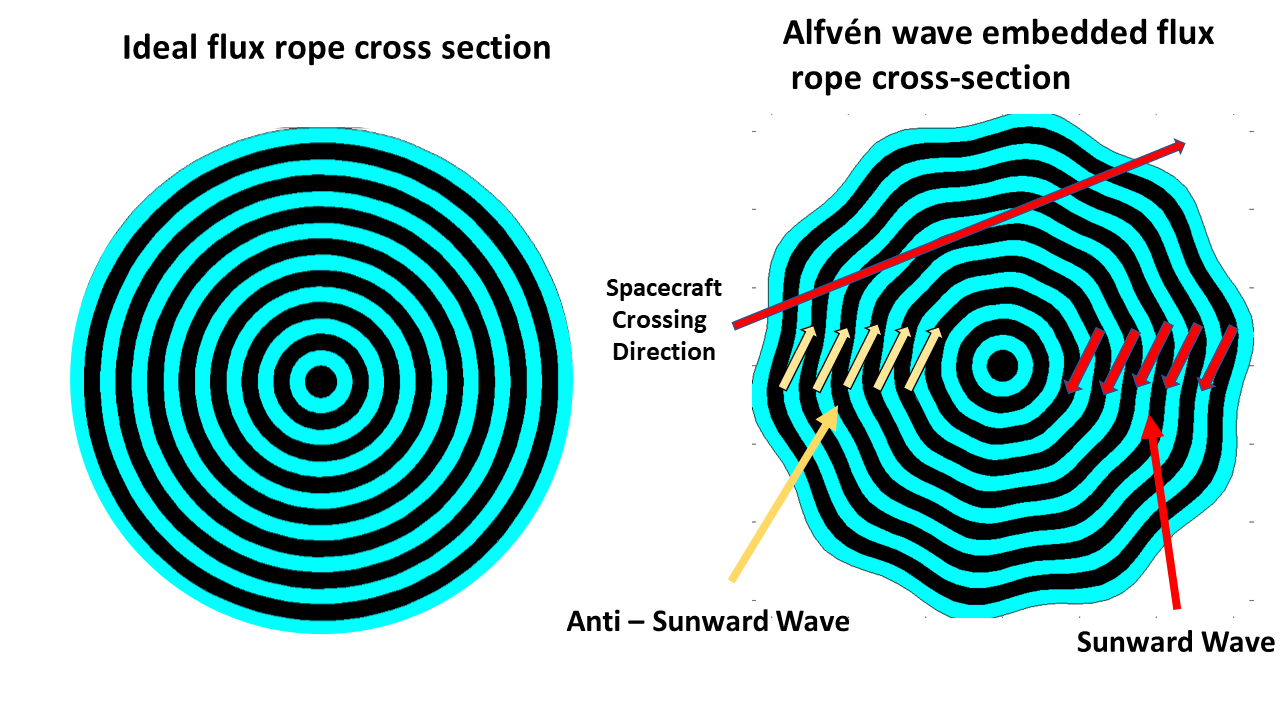}
	\caption{Cartoon demonstration of the generation of SAW, the first image is about the ideal flux rope cross-section, and the second cartoon image manifests the spacecraft crossing one of the cross sections of the flux rope.}
	\label{fig:cartoon}
\end{figure}

As an analogy, the bidirectional flow of Alfv\'en wave in the ICME flux rope can be explained as follows. Consider the cartoon picture of the cross-section of the flux rope, as shown in Figure \ref{fig:cartoon}. The left image is the ideal circular cross-section of the flux rope, whereas the right image demonstrates the SAW's superposed flux rope cross-section. The red arrow indicates the spacecraft's passage through the ICME flux rope. At the spacecraft's entry point, i.e., the anterior of the flux rope, the surface wave depicts the upward direction of propagation, whereas the posterior shows the downward movement of propagation. The upward and downward directions are proxies for wave propagation. An important fact is that the wave propagates inward, in the initial part of the MC and outward at the trailing region of the MC (see Figure \ref{fig:filter}). When the spacecraft moves from one end of the flux rope's cross-section to the other, the amplitude of fluctuations on the surface is more significant than at the centre. The outer layer of the flux rope describes higher amplitude of the Alfv\'enic oscillations. As we move into the inner concentric layers, the amplitude decreases and is minimum at the centre of the flux rope. This is also clearly seen in the top three panels of Figure \ref{fig:my_label}.


SAWs are expected to exist in astrophysical plasmas where density or magnetic field jumps occur. e.g., on the surfaces of magnetic flux tubes in the solar and stellar atmospheres, at the interfaces between plasmas of different properties in the solar wind and Earth's magnetosphere \citep{cramer2011physics}. We also observed a large proton temperature anisotropy at the front and trailing part of the flux rope, prompting us to believe that the large temperature anisotropy could be the possible source of SAW generation or vice-versa. Besides this, we observed the spike in number density ($N_p$) and sudden drop in the total IMF strength in the sheath region, just prior to the onset of the MC (see figure ~\ref{fig:IP_label}) which could be attributed to reconnection exhaust. Moreover, \cite{gosling2005direct} claims that AWs could be generated by reconnection exhaust at the quasi stationary heliospheric current sheet (HCS), implying that the difference in density variations at the boundary between the sheath region and MC leads to the induction of oppositely directed waves in the sheath region and the MC's front part. 

An AW is often observed in interplanetary space and most recently, in the flux ropes \citep{gosling2010torsional,raghav2018does}. The behaviour of a SAW is solely different from that of a classical AW \citep{goossens2012surface}. 
The SAWs are linked to tearing mode instability, leading to the time-dependent magnetic reconnection process. This results in the neutral collision effect and ionisation fraction to significantly impact the SAWs \citep{uberoi1994resonant,uberoi1996surface}. 
The resonant absorption of SAWs appears to be a viable heating process for both, the open regions and coronal loops \citep{ionson1978resonant}. Moreover, SAWs can damp through viscosity, resistivity and other kinetic factors, which results in the plasma wave heating \citep{evans2012coronal}. \cite{evans2009surface} explored the role of SAWs damping to the solar wind heating.  Here, a number of intriguing scientific questions arise, such as How does SAW alter the properties of the ICME? How do they dissipate? Additionally, how does the SAW amplitude change as the ICME moves across the heliosphere? These issues are beyond the purview of this article, although we might look into them later.


\section{acknowledgment}

Authors thanks to Mr. Greg Hilbert for valuable suggestion. The authors thank everyone involved in the WIND spacecraft mission development, including the data-providing team. We also appreciate NASA/Space GSFC's Physics Data Facilities (CDAWeb or ftp) service. SERB, India, is acknowledged since AR and OD are supported by SERB project reference file CRG/2020/002314. ZS also thanks the government of India's `The Department of Science and Technology (DST)' (urlhttps://dst.gov.in/) for their cooperation.

\bibliographystyle{unsrtnat}
\bibliography{ref}  






\end{document}